\documentclass[twocolumn,showpacs,preprintnumbers,amsmath,amssymb,floatfix]{revtex4-1}
\usepackage{graphicx}
\usepackage{color}

\begin{document}

\title{Contact effects on transport in magnetite, an archetypal correlated transition metal oxide}

\author{A.~A.~Fursina$^{1}$, R.~G.~S.~Sofin$^{2}$, I.~V.~Shvets$^{2}$, D.~Natelson$^{3, 4}$}

\affiliation{$^{1}$ Department of Chemistry, Rice University, 6100 Main St., Houston, TX 77005}
\affiliation{$^{2}$ CRANN, School of Physics, Trinity College, Dublin 2, Ireland}
\affiliation{$^{3}$ Department of Physics and Astronomy, Rice University, 6100 Main St., Houston, TX 77005}
\affiliation{$^{4}$ Department of Electrical and Computer Engineering, Rice University, 6100 Main St,.Houston, TX 77005}

\date{\today}


\begin{abstract}

Multiterminal measurements have typically been employed to examine electronic properties of strongly correlated electronic materials such as transition metal oxides without the influence of contact effects.   In contrast, in this work we investigate the interface properties of Fe$_3$O$_4$ with different metals, with the contact effects providing a window on the physics at work in the correlated oxide.  Contact resistances are determined by means of four-terminal electrical measurements as a function of source voltage and temperature.   Contact resistances vary systematically with the work function of the electrode metal, $\phi(M)$, $M=$Cu, Au and Pt, with higher work function yielding lower contact resistance.  This trend and the observation that contact resistances are directly proportional to the Fe$_3$O$_4$ resistivity are consistent with modeling the oxide as an effective $p$-type semiconductor with hopping transport.   The jumps in contact resistance values at the bias-driven insulator-metal transition have a similar trend with $\phi$($M$), consistent with the transition mechanism of charge gap closure by electric field.

\end{abstract}
 
\maketitle

Transition metal oxides (TMO) with strong 3$d$-electron correlations exhibit a rich variety of physical properties \cite{2003_e_correl_Tokura}, such as metal-insulator transitions, high temperature superconductivity, magnetic ordering, resistive switching, etc.  Bulk electronic transport of such strongly correlated electronic systems (SCESs) have been examined extensively.  However, there has been little study of the contacts between normal metals (M) and SCESs, even though injecting charge into a SCES involves ``dressing'' normal metal quasiparticles due to strong correlations.  M/SCES interfaces under bias have been considered theoretically, and recent studies suggest SCESs with correlation-induced gaps in the density of states may be treated as classical semiconductors with minor modifications\cite{2005_PRL_MTMO_interface}; correlation effects, however, are predicted to suppress rectification at M/SCES interface \cite{2007_PRB_suppr_rectific}.  In such gapped SCESs, breakdown of the insulating state upon application of an external electric field is theoretically described as analogous to Landau-Zener tunneling in band insulators\cite{2003_PRL_Mott_break}, which is effectively the closure of the charge gap by electric field\cite{2008_PRB_theory}.

Contacts also play a critical role in resistive switching (RS), in which a significant change of electrical resistance (from a comparatively insulating Off state to a more conductive On state) is induced by an application of external stimuli such as current, voltage or electric field\cite{2008_Sawa_review}.  Electric field driven breakdown of a gapped correlated state has been proposed as the mechanism for RS in complex manganites\cite{1997_Tokura_first} and magnetite\cite{Our_magnetite_2008,2009_PRB_hyster,2010_my_PRB2}.  Nonvolatile RS has also been observed in other TMOs, including perovskites\cite{2007_Waser}, TiO$_2$ \cite{2005_JAP_TiO2_Waser,Kwon_naturenano_2010}, and doped SrTiO$_3$ \cite{2006_Nature_SrTiO3_Waser}.  Nonvolatile RS systems are of much technological interest\cite{2009_Waser_review}.  Experiments probing the $M$/$TMO$ interface properties in nonvolatile RS systems are scarce.  Multilead electrical measurements\cite{2003_3lead_Baikalov} and impedance spectroscopy\cite{2009_bulk_inter_Waser} studies show that the RS takes place at the interface.  RS behavior is dependent on the work function of the metal in contact with the TMO \cite{2004_diff_cont_Me_Tokura,2009_JElChem_DCM2}; RS is observed only if the contact metal forms a rectifying contact with the oxide material\cite{2006_APL_DCM1,2008_Sawa_review}.

We present a systematic experimental study of interface transport properties between a SCES TMO, magnetite (Fe$_3$O$_4$), and three metals, $M=$Cu, Au, Pt, using nanostructure techniques.  Stoichiometric bulk magnetite undergoes a phase transition between a high temperature ``bad metal'' phase and a low temperature insulating phase at the Verwey temperature, $T_{\mathrm{V}} \approx$~122~K\cite{Walz2002}, and exhibits electric field driven RS below $T_{\mathrm{V}}$\cite{Our_magnetite_2008,2009_PRB_hyster,2010_my_PRB2}.
We use multiterminal measurements to assess contact resistances as a function of bias voltage, temperature, and work function, $\phi$, of the contacting metal $[\phi$(Cu)$<$$\phi$(Au)$<$$\phi$(Pt)$]$.  For all contacting metals, low bias contact resistances are found to track the bulk resistivity as a function of temperature, with larger $\phi$ correlating with lower contact resistance.  We also track contact resistances and their changes through the RS transition.  These data are consistent with field-driven gap closure as the RS mechanism.

\begin{figure}[h!]
\begin{center}
\includegraphics[clip,width=8.5cm]{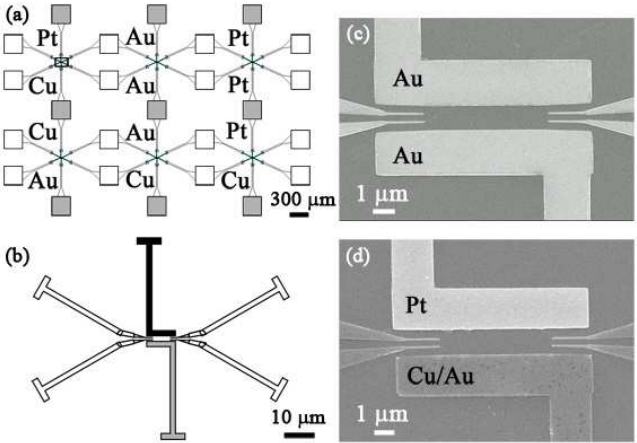}
\end{center}
\vspace{-5mm}
\caption{\small (a) NPGS pattern of different contact metal devices on the same piece of Fe$_3$O$_4$ film. Grey squares are 300~$\mu$m $\times$ 300~$\mu$m pads leading to source and drain electrodes, while pads colored in white lead to the voltage probes. The very small box in the center of the upper-left device defines a region magnified in (b).  Colored in black and grey, the $\lceil$- and $\rfloor$-shaped source/ drain leads are the ones made of different contact metals. (c) and (d) are SEM images of four-terminal devices with Au-Au and Cu/Au-Pt  source-drain leads, respectively.}
\label{fig1}
\vspace{-5mm}
\end{figure}

The 50~nm Fe$_{3}$O$_{4}$ (100) thin films used in the present study were grown on (100) oriented MgO single crystal substrates  as described elsewhere \cite{Shvets_backscat,Shvets_high_res_Xray}.   
Devices for four-terminal measurements were prepared on the surface of the Fe$_{3}$O$_{4}$ films by electron beam lithography and electron-beam deposition of contact metals, as described in detail in Ref.~\cite{2010_my_PRB2}. 
To rule out sample-to-sample variations, sets of devices were made incorporating different combinations of contact metals as source and drain electrodes: Cu-Au and Cu-Pt, as well as Au-Au and Pt-Pt, on a single magnetite wafer.  Thus all  devices underwent the same fabrication steps (Fig.~\ref{fig1}a), and all devices had the same geometrical characteristics.   The Verwey temperature of the resulting Fe$_{3}$O$_{4}$ devices, found from zero-bias resistance, $R$, as a function of temperature, $T$, is $\sim$95~K, well below the $T_{V}\approx$110~K of pristine films grown under identical conditions.  This difference is due to the intensive fabrication procedure, likely from deviations from the ideal Fe$_{3}$O$_{4}$ stoichiometry, known to suppress $T_{\mathrm{V}}$\cite{1985_PRB_nonstoich_2, 1985_PRB_nonstoich_1}.

The whole four-terminal pattern except for the $\lceil$- and $\rfloor$-shaped source/drain leads (see Fig.~\ref{fig1}b), are made with a 6~nm Cu adhesion layer\cite{2010_my_PRB2} and a 10~nm Au cover layer.  The $\lceil$ and $\rfloor$ source-drain leads are made of 20~nm layers of either Cu, Au or Pt (Fig.~\ref{fig1}b).  This design ensures as much as possible that all devices have identical voltage probes, and that the difference in performance is indeed due to the difference in the interface with the source and drain contact metal.   Some devices have shared large contact pads (Fig.~\ref{fig1}a). 

\begin{figure}[h!]
\begin{center}
\includegraphics[clip,width=8.5cm]{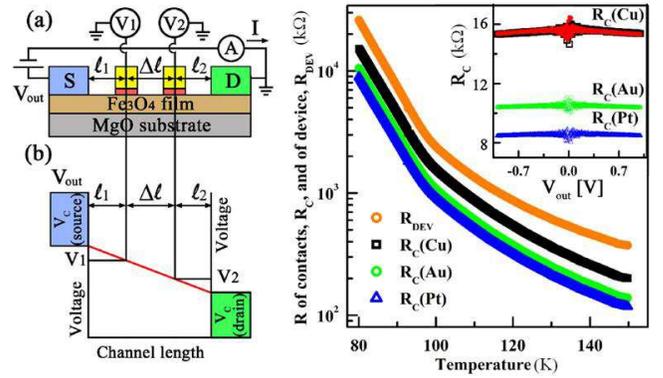}
\end{center}
\vspace{-5mm}
\caption{\small Schematics of (a) electrical circuit for four-probe measurements. Letters S and D  denote source and drain contacts, respectively, and (b) how the source voltage, $V_{\mathrm{out}}$, is distributed along the channel. (c) Temperature dependence of the low bias device resistance, $R_{\mathrm{DEV}}$, and contact resistance $R_{\mathrm{C}}$($M$), $M$=Cu, Au, Pt, demonstrating $R_{\mathrm{C}}$(Pt)$<R_{\mathrm{C}}$(Au)$<R_{\mathrm{C}}$(Cu), and that for all cases $R_{\mathrm{C}} \sim R_{\mathrm{DEV}}$.  Inset shows the weak dependence of $R_{\mathrm{C}}$($M$) on source voltage in the (nearly) linear regime around 0~V at 80~K.}
\label{fig2}
\vspace{-5mm}
\end{figure}

Fig.~\ref{fig1}a shows the device layout.  Representative SEM images of Au-Au and Cu/Au-Pt devices are shown in Fig.~\ref{fig1}c and d, respectively, demonstrating source and drain electrodes along with two pairs of voltage probes within the $\sim$1$\mu$m-long channel.   Gold wires (25~$\mu$m diameter) were attached by In soldering to the 300~$\mu$m $\times$ 300~$\mu$m pads of the devices to connect to the pins of the measurement puck.  The puck was placed into a cryostat (300~K - 80~K) for measurement.  Electrical characterization was performed via standard four-terminal methods using a semiconductor parameter analyzer (HP 4155A).  The voltage, $V_{\mathrm{out}}$, is applied to the source lead with the drain grounded, and current flowing through the channel is recorded, while a pair of voltage probes senses voltages $V_1$ and $V_2$ (and therefore $\Delta V \equiv V_{1}-V_{2}$), as shown in Fig.~\ref{fig2}a.  

The total applied potential difference, $V_{\mathrm{out}}$, must equal the sum of voltage drops across a device.  There is a voltage across the source electrode/Fe$_3$O$_4$ interface $[V_{\mathrm{C}}$(source)$]$ (Fig.~\ref{fig2}b).  In the assumption of a homogeneous film exhibiting ohmic conduction between the electrodes, voltage linearly drops across the channel.  Finally, there is a further voltage drop at the Fe$_3$O$_4$/drain electrode interface $[V_{\mathrm{C}}$(drain)$]$ ((Fig.~\ref{fig2}b)).  With these assumptions and using the geometrical characteristics of our devices as obtained from SEM images ({\it i.e.}, $\ell_1$, $\Delta \ell$, and $\ell_2$, see Fig.~\ref{fig1}c) we can calculate
$V_{\mathrm{C}}$(source) and $V_{\mathrm{C}}$(drain)\cite{2010_my_PRB2} at each $V_{\mathrm{out}}$.

Four measurements were conducted at each $T$ for each four-terminal device with two pairs of voltage probes.  With source and drain chosen, measurements were made with each pair of $V$ probes; source and drain were then reversed, and the $V$ probe four-terminal measurements were then repeated.  From the resulting data, $R_{\mathrm{C}}$(source), $R_{\mathrm{DEV}}$ and $R_{\mathrm{C}}$(drain) (the source contact resistance, the device resistance, and the drain contact resistance, respectively) were calculated at each $V_{\mathrm{out}}$ according to the procedure described below and in Ref.~\cite{2010_my_PRB2}.  Values obtained with different source/drain and voltage probe configurations were checked for consistency.  Calculations give the same values of $R_{\mathrm{DEV}}$, $R_{\mathrm{C}}$(source) and $R_{\mathrm{C}}$(drain) for left and right pairs of voltage probes, and $R_{\mathrm{C}}$(source) and $R_{\mathrm{C}}$(drain) switch appropriately upon exchanging the biasing/grounding choice.

We first consider the properties of the $M$/Fe$_{3}$O$_{4}$ interface near zero bias.  Around 0~V (typically -1~V to 1~V range of $V_{\mathrm{out}}$) current-voltage ($I$-$V_{\mathrm{out}}$) characteristics of the devices are linear.  In this Ohmic regime the contact resistances at each $V_{\mathrm{out}}$ can be calculated as $R_{\mathrm{C}}$(source)$\equiv V_{\mathrm{C}}$(source)/$I$ and $R_{\mathrm{C}}$(drain)$\equiv V_{\mathrm{C}}$(drain)/$I$, where $I$ is a measured current at a certain $V_{\mathrm{out}}$.  This data is shown as a function of temperature in Fig.~\ref{fig2}c for a particular source/drain and voltage probe configuration, measured with $|V_{\mathrm{out}}|<1$~V.  The Verwey transition is apparent as the kink and change in slope at 95~K.  The trend $R_{\mathrm{C}}$(Pt)$<R_{\mathrm{C}}$(Au)$<R_{\mathrm{C}}$(Cu) is consistent for all temperatures.  It is striking that $R_{\mathrm{C}}(T)$ tracks $R_{\mathrm{DEV}}(T)$, proportional to the magnetite resistivity, over the whole temperature range.  We discuss this further below. 

The inset of Fig.~\ref{fig2}c demonstrates the weak voltage dependence at low bias of $R_{\mathrm{C}}$ for Cu, Au and Pt at 80~K, extracted from the probing of Cu-Au and Cu-Pt devices.  The plots clearly illustrate that $R_{\mathrm{C}}$(Pt)$<R_{\mathrm{C}}$(Au)$<R_{\mathrm{C}}$(Cu). Note, that $R_{\mathrm{C}}$(Cu) extracted from the independent measurements of Cu-Au (closed red dots) and Cu-Pt (open black squares) devices is essentially identical.

The observed trend of $R_{\mathrm{C}}$ with metal type is consistent with the energetic offset between the Fermi level of the metal and the states responsible for transport in the magnetite.  In the absence of interfacial dipoles and Fermi level pinning, this offset is determined by the difference in the work function, $\phi$, of a contact metal $M$ and Fe$_3$O$_4$.  The work function of clean magnetite is $\phi_{bulk}$(Fe$_3$O$_4$)=5.78~eV, while $\phi$ of clean Cu, Au and Pt are 4.65~eV, 5.1~eV and Pt 5.65~eV, respectively\cite{work_func_list}.  While work functions can change due to surface termination and adsorbates\cite{Hild2008}, the systematic trend suggests that the ordering of work functions remains the same.  Magnetite surfaces were freshly cleaned immediately prior to insertion of the samples into the metal deposition system.  For our metals, all of the $M$/Fe$_3$O$_4$ contacts appear to have $\phi$(Fe$_3$O$_4$)$>\phi$($M$).   

The dominant type of charge carriers in Fe$_3$O$_4$ depends on the stoichiometry of the material as well as the temperature range\cite{fe3o4_p_type1,Fernandez2008}.  In bulk stochimetric magnetite, below $T_{\mathrm{V}}$ the charge carriers (as determined via Hall effect measurements) are holes.   Fe$_3$O$_4$ might therefore be modeled as a $p$-type semiconductor (s/c)\cite{fe3o4_p_type1,fe3o4_p_type2}, though it is important to keep in mind that the charge carriers in magnetite are not well-described by the picture of extended states in conventional crystalline semiconductors\cite{Walz2002}.  The hopping character of charge transport in magnetite both above and below the Verwey transition temperature suggests that a more apt comparison would be with highly disordered semiconductors, such as organic polymers\cite{Scott2003,Hamadani2004}.  In such hopping systems, when a carrier is injected from a metal electrode, there is a competition between diffusion of the carrier away from the contact by hopping, and the attractive interaction between the localized carrier and its image potential in the metal electrode.  As a result, lower bulk resistivity leads to lower effective contact resistance\cite{Scott2003}.

\begin{figure}[h!]
\begin{center}
\includegraphics[clip,width=5cm]{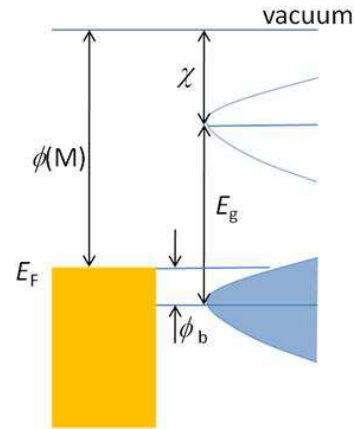}
\end{center}
\vspace{-5mm}
\caption{\small Schematic diagram of the suggested energetics at the metal/oxide interface below the Verwey temperature.  Various parameters are defined in the text.  Charge transport takes place between the single-particle states at the metal Fermi level and \textit{localized} states in the broadened ``valence'' band of the oxide.  Above the Verwey temperature, the situation is similar, with transport in the oxide still taking place via the hopping of localized carriers.  The trends in the contact resistance data suggest that in our samples dominant carriers are hole-like throughout the whole temperature range. }
\label{fig:banddiagram}
\vspace{-5mm}
\end{figure}

A schematic of the suggested interface energetics is shown in Fig.~\ref{fig:banddiagram}.  The key difference between this and the conventional Schottky case is that the electronic states in the broadened ``valence'' and ``conduction'' bands of the oxide are \textit{localized}, so that conduction takes place via hopping\cite{Walz2002}.  In the framework of conventional barrier formation at $M$/$p$-type s/c interface with $\phi$($p$-type s/c)$>\phi$($M$), the height of the Schottky barrier, $\phi _{b}$ is:
\begin{equation}
\phi_{b}=E_{g}-[\phi(M)-\chi]=E_{g}+\chi-\phi(M)
\label{SchotHeight}
\end{equation}
where $E_{g}$ and $\chi$ are a band gap and an electron affinity of a semiconductor, respectively\cite{semicond_book_Singh}.  According to Eq.~\ref{SchotHeight}, the higher $\phi$($M$) the lower the barrier height, $\phi_{b}$ and, thus, the lower $R_{\mathrm{C}}$ of the contact interface.  This is consistent with the observed trend in Fig.~\ref{fig2}c below $T_{\mathrm{V}}$.  The further observation that $R_{\mathrm{C}}(T) \propto R_{\mathrm{DEV}}(T)$ is consistent with charge injection into other hopping systems\cite{Scott2003,Hamadani2004}.  Note that this does not imply that there is no change in conduction across the Verwey transition.  Rather, it implies that transport in the oxide both above and below the transition temperature involves hopping of charge carriers, rather than transport via extended states.  

It is important to note that the effective bulk resistivity of magnetite films may be strongly affected by antiphase boundaries (APBs), unavoidable at some density in films due to the kinetics of nucleation and growth.  APBs are observed in Fe$_{3}$O$_{4}$ epitaxial films grown onto MgO (100) substrates\cite{Arora2006,Eerenstein2002}.  Since the Fe$_{3}$O$_{4}$ $(Fd{\overline{3}}m)$ crystal structure is of lower symmetry than MgO ($Fm3m$), there are several equivalent nucleation sites on the MgO (100) surface, which leads to the formation of APBs at the interface of neighboring grains.  The presence of APBs leads to a large difference in the physical properties of epitaxial magnetite films and those of the bulk, including an enhanced resistivity\cite{Eerenstein2002}, magnetoresistance\cite{Eerenstein2002PRL}, and magnetization that does not saturate at high magnetic fields\cite{Margulies1997}.  The complicated magnetic microstructures occurring due to the APB network is likely to reduce the spin polarization of magnetite films\cite{Kasama2006}, though spin polarization is not relevant to the present work.  In a careful study of the thickness dependence of magnetite film properties, Eerenstein \textit{et al.} observed\cite{Eerenstein2002} that APBs contribute significantly to effective resistivity at thicknesses below 30~nm.  In our 50~nm films, measured resistivity is comparable to bulk resistivity, suggesting that APBs are not the dominant source of scattering.  It would be very interesting to perform contact resistance measurements in very thin films (in the limit that APBs strongly modify the bulk resistivity), to see whether the trend shown in Fig.~\ref{fig2}c continues to hold; one would expect not.

\begin{figure}[h!]
\begin{center}
\includegraphics[clip,width=8.5cm]{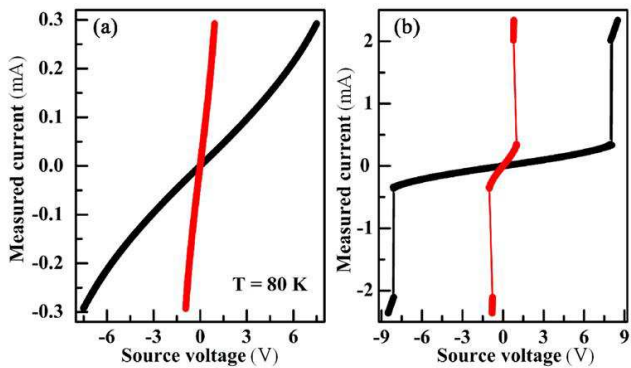}
\end{center}
\vspace{-5mm}
\caption{\small $I$-$V_{\mathrm{out}}$ (black) and $I$-$\Delta V$ (red) characteristics at 80~K (a) in non-linear regime before transition and (b) if applied voltage exceed a critical switching value, $V_{sw}$, inducing resistive switching.}
\label{IV_trans}
\vspace{-5mm}
\end{figure}

Now we consider $M$/Fe$_3$O$_4$ contacts at higher biases, far away from equilibrium.  As the applied voltage increases, $I$-$V_{\mathrm{out}}$ curves become non-linear, but remain symmetrical around 0~V (Fig.~\ref{IV_trans}a), in contrast to the rectifying curves of typical metal/semiconductor junctions.  At the same time, $I$-$\Delta V$ curves remain linear (Fig.~\ref{IV_trans}a) even at high biases, suggesting that the non-linearity in $I$-$V_{\mathrm{out}}$ originates from the $M$/Fe$_3$O$_4$ interfaces.  The absence of significant rectification is consistent with expectations from strong electron-electron correlations and is typical of TMOs \cite{2007_PRB_suppr_rectific}.  As $V_{\mathrm{out}}$ increases further, as reported previously\cite{Our_magnetite_2008,2009_PRB_hyster,2010_my_PRB2} there is a sharp jump in current as $V_{\mathrm{out}}$ reaches a critical switching value, $V_{\mathrm{sw}}$ (Fig.~\ref{IV_trans}b).  The system after this RS transition is characterized by an approximately linear $I$-$V_{\mathrm{out}}$ dependence and lower resistance compared to the state before transition.   RS in magnetite devices is observed for contacts made from a variety of low- and high-work function metals (4.65~eV - 5.65~eV range for Cu, Au and Pt, as well as Fe, Ti and Al, which have even lower $\phi$s of 4.5~eV, 4.33~eV and 4.28~eV \cite{work_func_list}, respectively). 

The observation of reproducible field-driven RS in magnetite across a wide variety of contact metals is consistent with the proposed field-driven gap closure mechanism, and stands in sharp contrast to trends seen in \textit{nonvolatile} RS systems.  In such systems, for which the origin of RS is attributed to the change in Schottky barrier height (or width) by trapped charge carriers at the interface states\cite{2007_Schottky_sol_solutions}, switching occurs only if the contact metal forms a rectifying contact with the oxide material.  That is, $p$-type materials (such as complex manganites $Re_{0.7}$A$_{0.3}$MnO$_3$ ($Re$ = Pr, La; $A$ = Ca, Sr)) demonstrate RS with low work-function metals, $M$=Mg, Al and Ti, and do not show RS with high work function metals such as Au and Pt \cite{2006_APL_DCM1}.  On the other hand, $n$-type materials (Nb or Cr doped SrTiO$_3$) demonstrate RS only when electrodes are made of high $\phi$ material, such as Au or SrTiO$_3$ \cite{2008_Sawa_review}.  The RS in magnetite, in contrast, is clearly a bulk rather than a contact phenomenon.

In the non-linear regime the definition of contact resistance as $R_{\mathrm{C}}\equiv V_{\mathrm{C}}$/$I$ is no longer physically meaningful.  In this case we estimate {\it differential} $R_{\mathrm{C}}$ as $\Delta V_{\mathrm{C}}/\Delta I$ continuously over narrow $\Delta V_{out}$ ranges ($\sim$ 0.5~V) in which $I$-$V_{out}$ curves are approximately linear.  In the apparently linear regime at biases above $V_{\mathrm{sw}}$, $R_{\mathrm{C}}$ is again calculated the same way as in linear regime around 0~V.

\begin{figure}[h!]
\begin{center}
\includegraphics[clip,width=8.5cm]{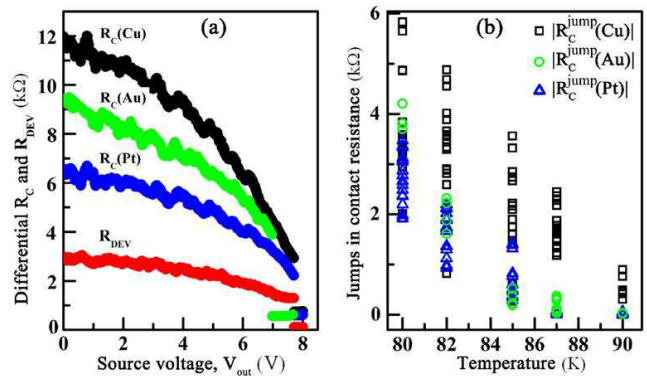}
\end{center}
\vspace{-5mm}
\caption{
\small (a) Calculated differential $R_{\mathrm{C}}$($M$), $M$=Cu, Au and Pt, and resistance of the magnetite channel, $R_{\mathrm{DEV}}$, at 82~K before and after transition point.
(b) Temperature dependence of the absolute values of jumps in contact resistances, $|R^{jump}_{\mathrm{C}}(M)|$. 
 Error bars represent a standard deviation over several independent measurements (different devices, left/right voltage pairs, different grounds, positive/negative switching branches).}
\label{fig4}
\vspace{-5mm}
\end{figure}

The representative dependences of calculated differential $R_{\mathrm{C}}$ on $V_{\mathrm{out}}$ for Cu, Au and Pt contact metals are shown in Fig.~\ref{fig4}a at 82~K (only positive voltage branch for clarity; negative branch is symmetrical). Reflecting the non-linearity in $I$-$V_{\mathrm{out}}$ curves, contact resistances decrease as $V_{\mathrm{out}}$ increases. The $R_{\mathrm{C}}$(Pt)$<R_{\mathrm{C}}$(Au)$<R_{\mathrm{C}}$(Cu) dependence is observed in the whole $V_{\mathrm{out}}$ range up to the transition point (Fig.~\ref{fig4}). 

On the onset of the transition, as $V_{\mathrm{out}}$ reaches critical $V_{\mathrm{sw}}$ value, contact resistances decrease abruptly and $R_{\mathrm{C}}$ values in the On state are approximately identical (about 600-700~$\Omega$ for this set of devices) for all three contact metals (Fig.~\ref{fig4}).  Note that contact resistances of {\it both} source and drain electrodes drop at the transition point.   The resistance of the magnetite channel, $R_{\mathrm{DEV}}$, also decreases at $V_{\mathrm{out}}$=$V_{\mathrm{sw}}$ (Fig.~\ref{fig4}a).  The absolute value of $R_{\mathrm{DEV}}$, however, still scales with the channel length (for details see Ref.\cite{2010_my_PRB2}), as expected.   It is clear from Fig.~\ref{fig4}a that $R_{\mathrm{DEV}}$ vs $V_{\mathrm{out}}$ dependence is ``flatter'' (closer to Ohmic), than the contact resistance's. This is a direct manifestation of the linearity of $I$-$\Delta V$ curves in contrast to the non-linearity of $I$-$V_{\mathrm{out}}$ curves (see Fig.~\ref{IV_trans}a).

In previous work\cite{2009_PRB_hyster,2010_my_PRB2} we proposed charge gap closure by electric field recently predicted in theory\cite{2003_PRL_Mott_break, 2008_PRB_theory} as consistent with RS in magnetite.  In this mechanism, with the assumption of Schottky-like barriers at $M$/Fe$_3$O$_4$ interfaces, the magnitudes of contact resistance jumps at a transition point, $R^{jump}_{\mathrm{C}}$, are expected to depend on the work function of the metal according to the relative alignment of Fermi level, $E_{\mathrm{F}}$, of the contact metal and effective $E_{\mathrm{F}}$ of Fe$_3$O$_4$.

The data presented experimentally demonstrate that $R^{Off}_{\mathrm{C}}$(Pt)$<R^{Off}_{\mathrm{C}}$(Au)$<R^{Off}_{\mathrm{C}}$(Cu) at all $V_{\mathrm{out}}$ before the transition, and $R^{On}_{\mathrm{C}}$(Pt)$\approx R^{On}_{\mathrm{C}}$(Au)$\approx R^{On}_{\mathrm{C}}$(Cu) after the transition (fig.~\ref{fig4}a).  Thus, jumps in $R^{jump}_{\mathrm{C}}$($M$)=$R^{On}_{\mathrm{C}}$($M$)$-R^{Off}_{\mathrm{C}}$($M$) are indeed $M$-dependent as $R^{jump}_{\mathrm{C}}$(Pt)$<R^{jump}_{\mathrm{C}}$(Au)$<R^{jump}_{\mathrm{C}}$(Cu). However, as can be seen in Fig.~\ref{fig4}a, the value of $R^{jump}_{\mathrm{C}}$ is also influenced by the $V_{\mathrm{sw}}$ value which, in turn, is proportional to the channel length and thus varies from device to device.  That is the reason that $R^{jump}_{\mathrm{C}}$ values vary significantly (see error bars in Fig.~\ref{fig4}b) in different measurements of different devices.  In some cases this variation obscures the trend $R^{jump}_{\mathrm{C}}$(Pt)$<R^{jump}_{\mathrm{C}}$(Au)$<R^{jump}_{\mathrm{C}}$(Cu), but overall the latter dependence stands true and consistent with expectations of the charge gap closure mechanism.

We have investigated $M$/Fe$_3$O$_4$ ($M$=Cu, Au and Pt) contacts in this correlated resistive switching system by means of four-terminal electrical measurements.  The dependence of contact resistances on metal work function $\phi$ is consistent with modeling the correlated oxide as an effective $p$-type semiconductor.  Likewise, the observed proportionality between contact resistances and bulk resistivity over a broad temperature range is consistent with models and observations of charge injection into semiconducting hopping systems.  We continuously traced the contact resistances as a function of bias before and after the resistive switching transition and showed that the jumps in contact resistances evolve with $\phi$ as $R^{jump}_{\mathrm{C}}$(Pt)$<R^{jump}_{\mathrm{C}}$(Au)$<R^{jump}_{\mathrm{C}}$(Cu) at all temperatures below $T_{V}$ when switching is observed.  This trend is consistent with the charge gap closure by electric field as a proposed driving mechanism of RS in magnetite.  These results demonstrate the utility of nanostructure contact studies in unraveling the physics at work in correlated materials.

This work was supported by the US Department of Energy grant DE-FG02-06ER46337.  DN also acknowledges the David and Lucille Packard Foundation and the Research Corporation.  RGSS and IVS acknowledge the Science Foundation of Ireland grant 06/IN.1/I91.



\begin{thebibliography}{10}

\bibitem{2003_e_correl_Tokura}
Y.~Tokura,
\newblock Phys. Today {\bf 56}, 50 (2003).

\bibitem{2005_PRL_MTMO_interface}
T.~Oka and N.~Nagaosa,
\newblock Phys. Rev. Lett. {\bf 95}, 266403 (2005).

\bibitem{2007_PRB_suppr_rectific}
K.~Yonemitsu, N.~Maeshima, and T.~Hasegawa,
\newblock Phys. Rev. B {\bf 76}, 235118 (2007).

\bibitem{2003_PRL_Mott_break}
T.~Oka, R.~Arita, and H.~Aoki,
\newblock Phys. Rev. Lett {\bf 91}, 066406 (2003).

\bibitem{2008_PRB_theory}
N.~Sugimoto, S.~Onoda, and N.~Nagaosa,
\newblock Phys. Rev. B {\bf 78}, 155104 (2008).

\bibitem{2008_Sawa_review}
A.~Sawa,
\newblock Mater. Today {\bf 11}, 28 (2008).

\bibitem{1997_Tokura_first}
A.~Asamitsu, Y.~Tomioka, H.~Kuwahara, and Y.~Tokura,
\newblock Nature {\bf 388}, 50 (1997).

\bibitem{Our_magnetite_2008}
S.~Lee, A.~Fursina, J.~T.~Mayo, C.~T.~Yavuz, V.~L.~Colvin, R.~G.~S.~Sofin, I.~V.~Shvets, and D.~Natelson,
\newblock Nature Mater. {\bf 7}, 130 (2008).

\bibitem{2009_PRB_hyster}
A.~A.~Fursina, R.~G.~S.~Sofin, I.~V.~Shvets, and D.~Natelson,
\newblock Phys. Rev. B {\bf 79}, 245131 (2009).

\bibitem{2010_my_PRB2}
A.~A.~Fursina, R.~G.~S.~Sofin, I.~V.~Shvets, and D.~Natelson,
\newblock Phys. Rev. B {\bf 81}, 045123 (2010).

\bibitem{2007_Waser}
R.~Waser and M. Aono,
\newblock Nature Mater. {\bf 6}, 833 (2007)


\bibitem{2005_JAP_TiO2_Waser}
B.~J. Choi, D.~S. Jeong, S.~K. Kim, C. Rohde, S. Choi, J.~H. Oh, H.~J. Kim, C.~S. Hwang, K. Szot, R. Waser, B. Reichenberg, and  S. Tiedke,
\newblock J. Appl. Phys. {\bf 98}, 033715 (2005).

\bibitem{Kwon_naturenano_2010}
D.-H. Kwon et al., \newblock Nature Nano. {\bf 5}, 148 (2010).

\bibitem{2006_Nature_SrTiO3_Waser}
K.~Szot, W.~Speier, G.~Bihlmayer, and R.~Waser,
\newblock Nat. Mater. {\bf 5}, 312 (2006).

\bibitem{2009_Waser_review}
R.~Waser,
\newblock Microelectron. Eng. {\bf 86} (2009).


\bibitem{2003_3lead_Baikalov}
A.~Baikalov, Y.~Q.~Wang, B.~Shen, B.~Lorenz, S.~Tsui, Y.~Y. Sun, Y.~Y.~Xu, and C.~W.~Chu,
\newblock Appl. Phys. Lett. {\bf 83}, 957 (2003).

\bibitem{2009_bulk_inter_Waser}
T.~Menke, P.~Meuffels, R.~Dittmann, K.~Szot, and R.~Waser,
\newblock J. Appl. Phys. {\bf 105} (2009).

\bibitem{2004_diff_cont_Me_Tokura}
A.~Sawa, T.~Fujii, M.~Kawasaki, and Y.~Tokura,
\newblock Appl. Phys. Lett. {\bf 85}, 4073 (2004).

\bibitem{2009_JElChem_DCM2}
M.~Hasan, R. Dong, H. Choi, J. Yoon, J.~Park, D.~Seong, H.~Hwang,
\newblock J. Electrochem. Soc. {\bf 156}, H239 (2009).

\bibitem{2006_APL_DCM1}
T.~Tokunaga, Y.~Kaneko, J.~P.~He, T.~Arima, A.~Sawa, T.~Fujii, M.~Kawasaki, and Y.~Tokura,
\newblock Appl. Phys. Lett. {\bf 88}, 223507 (2006).

\bibitem{Walz2002}
F. Walz, \newblock J. Phys.: Condens. Matter {\bf 14}, R285 (2002).

\bibitem{Shvets_backscat}
A.~Koblischka-Veneva, M.~R.~Koblischka, Y.~Zhou, S.~Murphy, F.~Mu\"{u}cklich, U.~Hartmann, and I.~V.~Shvets,
\newblock J. Magn. Magn. Mater. {\bf 316}, 663 (2007).

\bibitem{Shvets_high_res_Xray}
S.~K. Arora, R.~G.~S. Sofin, I.~V. Shvets, and M.~Luysberg,
\newblock J. Appl. Phys. {\bf 100}, 073908 (2006).

\bibitem{1985_PRB_nonstoich_2}
J.~P. Shepherd, R.~Arag\'on, J.~W. Koenitzer, and J.~M. Honig,
\newblock Phys. Rev. B {\bf 32}, 1818 (1985).

\bibitem{1985_PRB_nonstoich_1}
R.~Arag\'on, D.~J. Buttrey, J.~P. Shepherd, and J.~M. Honig,
\newblock Phys. Rev. B {\bf 31}, 430 (1985).

\bibitem{work_func_list}
H.~Michaelson,
\newblock J. Appl. Phys. {\bf 48}, 4729 (1977).

\bibitem{Hild2008}
K. Hild, J. Maul, T. Meng, M. Kallmayer, G. SCh{\"o}nhense, H. J. Elmers, R. Ramos, S. K. Arora, and I. V. Shvets.  \newblock J. Phys. Condens. Matter {\bf 20}, 235218 (2008).

\bibitem{fe3o4_p_type1}
V.~Shchennikov, S.~Ovsyannikov, A.~Karkin, S.~Todob, and Y.~Uwatokob,
\newblock Solid State Commun. {\bf 149}, 759 (2009).

\bibitem{Fernandez2008}
A.~Fern\'andez-Pacheco et al., \newblock Phys. Rev. B {\bf 77}, 100403(R) (2008).

\bibitem{fe3o4_p_type2}
D.~Kim and J.~M. Honig,
\newblock Phys. Rev. B {\bf 49}, 4438 (1994).

\bibitem{Scott2003}
J.~C. Scott, \newblock J. Vac. Sci. Technol. A {\bf 21}, 521 (2003).

\bibitem{Hamadani2004}
B.~H. Hamadani and D. Natelson, \newblock Appl. Phys. Lett. {\bf 84}, 443 (2004).

\bibitem{semicond_book_Singh}
J.~Singh,
\newblock {\em Semiconductor Devices: Basic Principles},
\newblock John-Wiley, 2001.

\bibitem{Arora2006}
S.~K. Arora, R.~G.~S. Sofin, I.~V. Shvets, and M. Luysberg.  \newblock J. Appl. Phys. {\bf 100}, 073908 (2006).

\bibitem{Eerenstein2002}
W.~Eerenstein, T.~T.~M. Palstra, T.~Hibma, and S.~Celotto.  \newblock Phys. Rev. B {\bf 66}, 201101(R) (2002).

\bibitem{Eerenstein2002PRL}
W.~Eerenstein, T.~T.~M. Palstra, S.~S. Saxena, and T.~Hibma.  \newblock Phys. Rev. Lett. {\bf 88}, 247204 (2002).

\bibitem{Margulies1997}
D.~T. Margulies, F.~T. Parker, M.~L. Rudee, F.~E. Spada, J.~N. Chapman, P.~R. Aitchison, and A.~E. Berkowitz. \newblock  Phys. Rev. Lett. {\bf 79}, 5162 (1997).

\bibitem{Kasama2006}
T.~Kasama, R.~E. Dunin-Borkowski, and W.~Eerenstein.  \newblock Phys. Rev. B {\bf 73}, 104432 (2006).

\bibitem{2007_Schottky_sol_solutions}
T.~Fujii, M.~Kawasaki, A. Sawa, Y. Kawazoe, H. Akoh, and Y.~Tokura,
\newblock Phys. Rev. B {\bf 75}, 165101 (2007).

\end{thebibliography}

\end{document}